\documentstyle{article}
\begin{document}
\centerline{\large\bf Role of causality in ensuring unconditional
security}
\vskip 1mm
\centerline{\large\bf of relativistic quantum cryptography}

\centerline{\small S.N.Molotkov and S.S.Nazin}

\centerline{\sl\small Institute of Solid State Physics of Russian
Academy of Sciences}

\centerline{\sl\small Chernogolovka, Moscow District, 142432 Russia}
\vskip 1mm

\begin{abstract}
The problem of unconditional security of quantum cryptography (i.e. the
security which is guaranteed by the fundamental laws of nature rather
than by technical limitations) is one of the central points in quantum
information theory. We propose a relativistic quantum cryptosystem and
prove its unconditional security against any eavesdropping attempts.
Relativistic causality arguments allow to demonstrate the security of the system in a simple way. Since the proposed
protocol does not employ collective measurements and quantum codes, the cryptosystem can be experimentally realized with the
present state-of-art in fiber optics technologies. The proposed cryptosystem employs only the individual measurements and
classical codes and, in addition, the key distribution problem allows to postpone the choice of the state encoding scheme
until after the states are already received instead of choosing it before
sending the states into the communication channel (i.e. to employ a sort of ``antedate'' coding). 
\end{abstract}


The idea of quantum cryptography was first proposed in the paper
of Wiesner [1] which, because of the novelty of the developed approach,
had not been published for a long time and only existed as a manuscript.
The idea of Wiesner became commonly known after the publication of the
paper of Bennett and Brassard [2]. An important advance was made in
Refs. [3] and [4]. Ekert proposed a cryptosystem based on the EPR-effect
[5]. Bennett {\it et al} [4] demonstrated that any eavesdropping attempt can be reliably detected employing an arbitrary
pair of non-orthogonal states. Later a large number of quantum cryptosystems and their realizations were proposed [6] which
are not guaranteed to be unconditionally secure. Currently there exist three proofs of the unconditional security. The proof
due to Mayers [7] addresses the so-called BB84 protocol [2] and so does the paper of Biham {\it et al} [8]. The proof of Lo
and Chau [9] deals with the protocol based on the EPR-effect [3] and, in contrast to Ref. [7], requires the access of
legitimate users to a quantum computer. The fact that these proofs have not yet become commonly accepted seems to be due to
the lack of a qualitative interpretation of their internal structure. Shor and Preskill [10] attempted to simplify these
proofs
by explicitly using the quantum codes. The first relativistic quantum
cryptosystem was proposed by Goldenberg and Vaidman [11]. The proof of
its
unconditional security is outlined in Ref. [12]. In our opinion, the
major obstacle in proving the unconditional security in Refs. [7--10]
arises because the corresponding protocols are formulated as the
exchange
protocols in the Hilbert state space and do not {\it explicitly} employ
the causality considerations and the fact that the legitimate users are
spatially separated although in real life the transmission of
information
always implies the preparation of information carriers (quantum
systems),
their propagation through the communication channel between two distant
users,
and, finally, performing a measurement of the quantum state of the
information
carriers at some later time. The restrictions imposed on the
measurability
of quantum states by special relativity were first considered by Landau
and
Peierls as early as in 1931 [13]. Further analysis was performed in the
paper of Bohr and Rosenfeld [14].

Let us begin with the formulation of the protocol security criterion.
Such a criterion seems to have been first explicitly formulated
in the work of Mayers [7]. Here we shall adopt a different criterion
which is more convenient for our proof. The protocol should satisfy
the two requirements which informally can be outlined in the following
way: the two strings of $N$ classical bits $s_A(N)$ and $s_B(N)$,
possessed by users A and B after the protocol is completed should be
(1) identical and (2) known to nobody else. More formally, a protocol
is secure if for {\it any $N\ge 1$ and any pair of real numbers
$\varepsilon_1>0$ and $\varepsilon_2>0$} its parameters
(the employed states, measurements, etc.) can be chosen in such a way
that:

\noindent
1. The probability for the two strings $s_A(N)$ and $s_B(N)$ to differ
in at least one bit is less than $\varepsilon_1>0$, i.e.
\begin{equation} 
\mbox{Pr}\{s_A(N) \neq s_B(N)\}\le \mbox{ } \varepsilon_1;
\end{equation}
in other words (in the language of mutual information between users A
and B)
for any $\varepsilon_1'>0$ it is possible to satisfy the inequality
\begin{equation}
I(A;B)\ge N-\varepsilon_1'.
\end{equation}

\noindent
2. The probability for the eavesdropper E to learn the string
$s_A(N)$ exceeds the probability of a simple guess, $2^{-N}$,
(remember that error probability associated with simply guessing a
particular bit
is 1/2 and represents the worst eavesdropper strategy)
by no more than $\varepsilon_2$:
\begin{equation}
\mbox{Pr}\{s_A(N)=s_E(N)\}\le 2^{-N} + \varepsilon_2;
\end{equation}
equivalently, the eavesdropper has arbitrarily small information on the
strings
$s_A(N)$ and $s_B(N)$ adopted as the key of length $N$ by legitimate
users:
\begin{equation} 
I(A;E)\le \varepsilon_2, \quad I(B;E)\le \varepsilon_2.
\end{equation}
Here the string of bits adopted as the key should not be understood
the bits the original bits sent by user A: each bit of the key is
actually
a function of the original bits sent by user A which passed the
appropriate
tests performed by user B aimed at detection of eavesdropping attempts.

If a single classical bit is to be transmitted, the user A associates
with
the logical states 0 and 1 of the classical bit the density matrices
$\rho_0$ and $\rho_1$ which can be chosen with the {\it a priori}
probabilities $\pi_0$ and $\pi_1$,  $\pi_0+\pi_1=1$.
Classical information is extracted by performing quantum measurements
on the system described by the density matrix
$\rho=\pi_0\rho_0+\pi_1\rho_1$.
The measurements are described by the identity resolutions in the
state space, $\sum_i E_i=I$. Information on the classical bit sent
by user A available to user B is defined as the mutual information
maximized over all possible measurements which can be performed by user
B:
\begin{equation}
I(A;B,\rho_0;\rho_1)=\max_{\{E_i\}}
\sum_i\left\{
\pi_0 \mbox{Tr}\{\rho_0E_i\}
\mbox{log}_2\left(\frac{\mbox{Tr}\{\rho_0E_i\} }{\mbox{Tr}\{\rho
E_i\}}\right)+
\pi_1 \mbox{Tr}\{\rho_1E_i\}
\mbox{log}_2\left(\frac{\mbox{Tr}\{\rho_1E_i\} }{\mbox{Tr}\{\rho
E_i\}}\right)
\right\}.
\end{equation}
A fundamental upper boundary on the available information is given by
the inequality first proved by Holevo [15] (see also [16]):
\begin{equation}
I(A;B,\rho_0;\rho_1)\le
S_{vN}(\rho)-\sum_{i=0,1}\pi_iS_{vN}(\rho_i),\quad
S_{vN}(\rho)=-\mbox{Tr}\{\rho\mbox{log}(\rho)\},
\end{equation}
where $S_{vN}(\rho)$ is the von Neumann entropy [17], and the equality
is
reached if and only if the density matrices $\rho_0$ and $\rho_1$
commute
whit each other. For pure states the latter means that the equality in
(6) is reached only for orthogonal states
($\rho_{0,1}=|\psi_{0,1}\rangle\langle\psi_{0,1}|$
and $\langle\psi_0|\psi_1\rangle=0$). In that case the available
information
reaches the maximum possible value of
\begin{equation}
I^{\max}(A;B,\rho_0;\rho_1)=1\quad E_0={\cal
P}_0=|\psi_0\rangle\langle\psi_0|,
\quad E_1={\cal P}_1=|\psi_1\rangle\langle\psi_1|.
\end{equation}
In other words, reliable distinguishability of the orthogonal quantum
states
allows to transmit the maximum possible classical information. However,
just
because of their reliable distinguishability they cannot be used in
quantum
cryptography (at least in the protocols employing the Hilbert state
space properties only).

It should be emphasized that it is implicitly assumed in Eqs. (5--7)
that
the entire Hilbert state space is always available for measurements. The
fact that the causality arguments are not explicitly used should be
understood as the possibility for user B to perform the measurement over
a quantum state $|\psi_i\rangle$ prepared by user A at time $t_A$ at
arbitrary
later time (formally, even at $t_B = t_A+0$). The measuring operators
$E_i$ at different times are related by the expression
$E_i(t_B)=U(t_B-t_A)E_iU^{-1}(t_B-t_A)$, where $E_i$ is the measuring
operator
taken at time $t_A$. Then the available information (5) does not depend
on time
and can be obtained immediately after the measurement is performed.

The above formulation of the problem is not only unnatural, but it does
not
correspond to the real procedure of information transmission between
two distant parties. As a matter of fact, the information (to be more
precise,
the physical quantum objects carrying that information) always propagate
from one user to the other. Therefore, it is much more natural to
formulate
the problem explicitly taking into account the causal relations between
the
states preparation, propagation, and measurements preformed on these
states
as they reach the second user and become available to his measurement
apparatus.

Since the information is carried in the Minkowskii space-time by real
physical
objects, e.g. photons (rather than the abstract physical systems and the
rays
in Hilbert state space ascribed to them as frequently assumed in the
non-relativistic quantum information theory), the role of the instrument
(superoperator) in the quantum field theory is played by the quantum
electrodynamical $\hat{\cal S}$-matrix.  The latter should satisfy
the unitarity and causality requirements first explicitly derived by
Bogolubov [18]:
\begin{equation}
\hat{\cal S}\hat{\cal S}^+=I,\quad
\frac{\delta \hat{\cal S}(g)}{\delta g(\hat{y})}\hat{\cal S}^+(g)=0,
\quad\mbox{for}\quad \hat{x}\le \hat{y},
\end{equation}
which means that the $\hat{\cal S}$-matrix does not depend on the
behavior
of $g(\hat{x})$ (the function specifying how the interaction is switched
on)
at point $\hat{x}\le \hat{y}$ separated from $\hat{y}$ by a space-like
interval. We do not know whether it is possible to develop a protocol
based on the first principles (general structure of the $\hat{\cal
S}$-matrix)
alone. Therefore, in the rest of the paper we shall adopt a simple
one-dimensional model containing all the necessary restrictions imposed
by the
relativistic causality. In addition, the adopted approach is further
justified
by the fact that the real fiber optics quantum communication channels
are
actually quasi-one-dimensional systems.

We shall first describe the states and measurements used in the
protocol.
Legitimate users control the spatially separated domains
$\Omega_A$ and $\Omega_B$ of size $L$. When the protocol is started at
$t_A=0$,
user A prepares with equal probabilities one of the following two
orthogonal
states corresponding to 0 or 1:
\begin{equation}
|\psi_{0,1}\rangle=\int_{0}^{\infty}dk {\cal
F}(k)a^{+}_{0,1}(k)|0\rangle=
\int_{0}^{\infty}dk {\cal F}(k)|k,e_{0,1}\rangle=|{\cal
F},e_{0,1}\rangle,
|k,e_{0,1}\rangle=a^{+}_{0,1}(k)|0\rangle, \quad
\langle k,e_{i}|k',e_j\rangle=\delta(k-k')\delta_{ij},
\end{equation}
where $a^{+}_{0,1}(k)$ is the operator creating a photon with momentum
$k>0$
and one of the two orthogonal polarization states $e_0$ and $e_1$,
${\cal F}(k)$ is the state amplitude in momentum representation,
$i,j=0,1,k\in(0,\infty)$.
In the position representation the states are written as
\begin{equation}
|\psi_{0,1}\rangle=\int_{-\infty}^{\infty}
{\cal F}(x-t)|x,t\rangle\otimes|e_{0,1}\rangle,\quad
{\cal F}(x-t)=\int_{0}^{\infty}dk {\cal F}(k)\mbox{e}^{ik(x-t)},\quad
\langle k|x,t\rangle=\frac{1}{\sqrt{2\pi}}\mbox{e}^{ik(x-t)},\quad
x,t\in(-\infty,\infty).
\end{equation}
The amplitude ${\cal F}(x-t)$ depends on the difference $x-t$ only,
in agreement with the intuitive picture of a packet propagating in the
positive direction of the $x$-axis with the speed of light and having
the spatio-temporal shape described by ${\cal F}(x-t)$.

It should be noted that the basis vectors $|x,t\rangle$ are not
orthogonal.
Normalization of the state vector in the position representation can be
written as [19]
\begin{equation}
\int_{-\infty}^{\infty}dx\mbox{e}^{ik(x-t)}\frac{1}{x-t+a}=
i\pi\mbox{ }\mbox{sgn}(k)\mbox{e}^{-ika},
\end{equation}
\begin{equation}
\langle\psi_{0,1}|\psi_{0,1}\rangle=\langle{\cal F}|{\cal F}\rangle=
\int_{-\infty}^{\infty} \int_{-\infty}^{\infty} dx dx'
{\cal F}(x-t){\cal F}^*(x'-t)[\frac{1}{2}\delta(x-x')+\frac{i}{\pi}
\frac{1}{x-x'}]= \int_{-\infty}^{\infty}|{\cal F}(x-t)|^2dx.
\end{equation}
The states are chosen to be almost ``monochromatic'', so that the
amplitude
${\cal F}(\tau)\approx const \approx 1/\sqrt{L}$ is actually represented
by a
constant wide ``plateau'' (to within the tails at its ends) and
\begin{equation}
\int_{\{L\}}dx |{\cal F}(x-t)|^2=1-\delta,\quad \delta\rightarrow 0.
\end{equation}
The decay at the ends can be chosen to be arbitrarily sharp and making
$\delta$ arbitrarily small. We shall assume that the latter condition is
satisfied and the parameter $\delta$ is well the smallest parameter in
the
problem\footnote{In our one-dimensional model the non-localizability
[20--22]
can be derived from the Wiener-Paley theorem [23]. Normalization and
square integrability conditions in the $k$-representation taken together
impose restrictions on the asymptotic behavior of the function 
${\cal F}(\tau)$:
$$ {\cal F}(\tau)=\int_{0}^{\infty} {\cal F}(k)\mbox{e}^{-ik\tau}dk,
\quad
\int_{-\infty}^{\infty}\frac{\textstyle |\mbox{ln}|{\cal F}(\tau)||}
{\textstyle 1+\tau^2}d\tau <\infty. $$ The function ${\cal F}(\tau)$
cannot
have a compact support with respect to $\tau$ and cannot decay
exponentially;
however, it can be arbitrarily strongly localized and possess a decay
rate
arbitrarily close to the exponential one, e.g. $$ {\cal F}(\tau)\propto
\exp{\{-\alpha\tau/\mbox{ln(ln...ln} \tau)\}}, $$ where $\alpha$ can be
any
real number. With this in mind, we shall for simplicity use the finite
domains when specifying the limits of integration.}.

Preparation of the extended states when the protocol is started at time
$t_A=0$ requires the access to the entire domain $\Omega_A$ of size $L$.
Intuitively, one can imagine a non-local device of size $L$ which
is simultaneously switched on at all point of the domain. There are no
any
formal arguments prohibiting such a state preparation procedure at time
$t_A=0$.
An extended state can also be prepared by a localized (point-like)
source
which is switched on at $t_A=0$ and produces (emits) a state propagating
into
the communication channel. For definiteness, it will be more convenient
for us
to assume that the state is prepared by a non-local source at time
$t_A=0$ and is immediately allowed at time  $t = t_A+0$ to propagate
as a whole into the communication channel.

{\it It is important for the protocol that the length of the quantum
communication channel $L_{ch}$ should not exceed the effective state
extent $L$, $L_{ch}<L$.}

At some later time $t_B$ when the entire state reaches the domain
$\Omega_B$
of size $L$ controlled by user B, he performs a measurement described
by the following identity resolution:
\begin{equation}
I=\int_{-\infty}^{\infty}dx |x,t_B\rangle\langle x,t_B|\otimes I_{{\bf
C}^2}=
{\cal P}_0(t_B)+{\cal P}_1(t_B)+ {\cal P}_{\bot}(t_B),\quad
{\cal P}_{0,1}(t_B)=|{\cal F}_{t_B},e_{0,1}\rangle
\langle {\cal F}_{t_B},e_{0,1}|,\quad
\end{equation}
\begin{equation}
|{\cal F}_{t_B},e_{0,1}\rangle=
\int_{-\infty}^{\infty}dx{\cal
F}(x-t_B)|x,t_B\rangle\otimes|e_{0,1}\rangle,
\quad x\in\Omega_B,\quad
{\cal P}_{\bot}(t_B)=I-{\cal P}_{0}(t_B)-{\cal P}_{1}(t_B).
\end{equation}
In other words, user B takes the projection on one of the two states
whose amplitude resides entirely in the domain $\Omega_B$. The
probabilities of different outcomes at time $t_B$ are
\begin{equation}
\Pr\{i,t_B;j\}=\mbox{Tr}\{|\psi_{i}\rangle\langle\psi_{i}|{\cal
P}_j(t_B)\}=
\delta_{ij} \int_{\{L\}} dx|{\cal F}(x-t_B)|^2=\delta_{ij},
\quad x\in\Omega_B.
\end{equation}

If the eavesdropper does not delay the states, he never has access to
them as
a whole. Therefore, the probability of incorrect state identification
by the eavesdropper will be different from zero even for orthogonal
states.
It should be emphasized that due to the orthogonal polarizations
our states are even locally orthogonal. Formally, that means that if at
certain
moment of time the eavesdropper has only access to a spatial domain
smaller than
the effective extent of the states to be distinguished, the state
identification
error probability is different from zero. The total error probability is
the
sum of two terms. The first one describes the situation when the
measuring
apparatus used by the eavesdropper did not fire at all. These outcomes
are
inevitable if the entire states are not available as a whole. The error
probability in that case is 1/2 (it is actually the error probability
for
simple guessing strategy). The second term in the total error
corresponds
to the case when the eavesdropper's apparatus fired in the spatial
domain
available to the eavesdropper. The state identification error in that
case
is strictly zero because of the local orthogonality of the states
employed.
Therefore, the total error is the product of the error for the case when
the
outcome occurred in the domain unavailable to the eavesdropper and the
fraction
of such outcomes. More formally, the total error is
\begin{equation}
P_e(t_E)=P_e(\Omega_E,t_E)+P_e(\overline{\Omega}_E,t_E);
\end{equation}
here $\Omega_E$  is the domain available to the eavesdropper
(accordingly,
$\overline{\Omega}_E$ is the unavailable domain, i.e. the completion of
$\Omega_E$ to the entire position space) and $t_E$ is the moment of time
when the measurement in the domain $\Omega_E$ was performed.
The total identity resolution is
\begin{equation}
I=I(\Omega_E,t_E)+I(\overline{\Omega}_E,t_E),\quad
I(\overline{\Omega}_E,t_E)=\sum_{i=0,1}
\int_{\overline{\Omega}_E}dx |x,t_E,e_i\rangle\langle x,t_E,e_i|,\quad
I(\Omega_E,t_E)=\sum_{i=0,1}
\int_{\Omega_B} dx |x,t_E,e_i\rangle\langle x,t_E,e_i|.
\end{equation}

Arbitrary strategy with a binary decision function for the outcomes in
the
domain $\Omega_E$ is described by an appropriate identity resolution on
$\Omega_E$. The lowest error probability $P_e(\Omega_E,t_E)$ is found by
the minimization over all possible identity resolutions $I(\Omega_E)$
[16]
\begin{equation}
P_e(\Omega_E,t_E)=\min_{E_0,E_1}\left\{
\frac{1}{2}\mbox{Tr}\{|\psi_0\rangle\langle\psi_0|E_1\}+
\frac{1}{2}\mbox{Tr}\{|\psi_1\rangle\langle\psi_1|E_0\}\right\}.
\end{equation}
$E_{0,1}$ is easily found and the total state identification error
$P_e(\Omega_E,t_E)\equiv 0$ proves to be
\begin{equation}
E_{0,1}=\int_{\Omega_E}dx |x,t_E;e_{0,1}\rangle\langle
x,t_E;e_{0,1}|,\quad
P_e(\overline{\Omega}_Et_E)=\frac{1}{2}
N(\overline{\Omega_E},t_E)=\frac{1}{2}
\int_{\overline{\Omega}_E} dx |{\cal F}(x-t_E)|^2.
\end{equation}
Accordingly, if the spatial domain $\Omega_E$ available to the
eavesdropper
has a fixed size, the probability of correct identification by the
eavesdropper
of the bit sent by user A is
\begin{equation}
P_{OK}(t_E)=1-P_e(\overline{\Omega}_Et_E)-P_e(\Omega_Et_E)=
\frac{1}{2}\left(1+\int_{\Omega_E} dx |{\cal F}(x-t_E)|^2\right).
\end{equation}
Thus the information on the bit sent by user A available to the
eavesdropper
depends on the size of the available domain and the choice of the moment
when
the measurement is performed. This information can be calculated using
Eq. (5)
taking into account that the measurement is described by the identity
resolution $\{ E_i \}=\{ I_{\overline{\Omega}_E}, E_0,E_1 \}$
(where $E_{0,1}$ are taken from Eq. (20)). The available information is
a sum
of two terms. The first term describes the part of the mutual
information
given by the outcomes in the unavailable domain (when the eavesdropper's
apparatus
did not fire at all) while the second one originates from the outcomes
in the
available domain $\Omega_E$:
\begin{equation}
I(A;E,\Omega_E,t_E)=I(A;E,\rho_0,\rho_1,\Omega_E,t_E)+
I(A;E,\rho_0,\rho_1,\overline{\Omega}_E,t_E).
\end{equation}
Calculation according to Eq. (5) taking into account that
$\{ E_i \}=\{ I_{\overline{\Omega}_E},   
E_0,E_1 \}$ and $\pi_0=\pi_1=1/2$ yields
\begin{equation}
I(A;E,\rho_0,\rho_1,\overline{\Omega}_E,t_E)=0,\quad
\mbox{Tr}\{\rho_{0,1}I_{\overline{\Omega}_E}\}=
\frac{1}{2}\mbox{Tr}\{\rho I_{\overline{\Omega}_E}\},\quad
\rho=\frac{1}{2}(\rho_0+\rho_1),
\end{equation}
\begin{equation}
I(A;E,\rho_0,\rho_1,\Omega_E,t_E)=\int_{\Omega_E} dx |{\cal
F}(x-t_E)|^2,\quad
\mbox{Tr}\{ \rho_{0}E_0 \}=\mbox{Tr}\{ \rho_{1}E_1 \}=
\mbox{Tr}\{\rho E_{0,1}\}.
\end{equation}
Therefore, the largest mutual information on the transmitted bit
available
to the eavesdropper depends on the moment of the measurement and the
fraction
of the state residing in the available domain. {\it In this way the
propagation
of information in the space-time is explicitly accounted for, in
contrast
to the protocols formulated in the state space only.} It is not
surprising
that the mutual information due to outcomes in the unavailable domain is
zero since the state identification error probability in that case is
1/2.
On the other hand, if the outcome took place in the available domain,
the identification error is zero, so that the mutual information is
proportional to the fraction of outcomes occurring in the available
domain.
To increase the mutual information, the eavesdropper should effectively
extend the available domain (wait until a larger part of the state
travel from
the domain controlled by user A to the domain available to his
measurements).

Let the effective increase in the domain size available to the
eavesdropper
(compared to the length of the communication channel which is supposed
to be
entirely available to him) is $\chi$. The correct identification
probability
for the eavesdropper is
\begin{equation}
\mbox{Pr}_E\{\chi \}=\frac{1}{2}\left(
1+\int_{\{ L_{ch}+\chi\} } dx |{\cal F}(x-t_E)|^2 \right)=
\frac{1}{2}\left(1+\frac{L_{ch}+\chi}{L}\right).
\end{equation}
For any state $|\tilde{\cal F}\rangle$ delayed by time (distance)
$\chi$,
the probability of passing a test for possible delay based on the
outcomes
of the measurement (14,15) performed by the legitimate user B is
\begin{equation}
\mbox{Pr}_B\{\chi\}=\mbox{Tr}\{ |\tilde{\cal F}\rangle\langle\tilde{\cal
F}|
({\cal P}_0(t_B)+{\cal P}_1(t_B)) \}=
\end{equation}
\begin{displaymath}
\Big|\int_{ \{ L-\chi \} } dx {\cal F}(x-t_B)\tilde{\cal
F}^*(x-t_B)\Big|^2\le
\left( \int_{\{L-\chi  \} } dx |{\cal F}(x-t_B)|^2 \right)
\left( \int_{\{L-\chi  \} } dx |\tilde{\cal F}(x-t_B)|^2 \right)\le
\left(1-\frac{\chi}{L}\right).
\end{displaymath}
It is sufficient to consider pure delayed states $\tilde{\cal F}$ only,
since
the linearity arguments can be applied to the case of mixed states. In
Eq. (26)
we took advantage of the Cauchy-Bunyakowskii inequality. The
integration domain is restricted to $L-\chi$ since because of the
limited
propagation velocity, no state delayed by time $\chi$ cannot reach by
the
moment of measurement $t_B$ the extreme right boundary of domain $L$;
remember also that ${\cal F}(x-t_B)=1/\sqrt{L}$.

Thus the probability for the eavesdropper to know the transmitted bit
and
simultaneously pass the test performed by user B is
\begin{equation}
\Pr\{bit_E=bit_A\bigwedge E\mbox{ }pass\mbox{ }test, \chi \}=
\mbox{Pr}_E\{\chi \}\cdot\mbox{Pr}_B\{\chi\}=
\frac{1}{2}\left(1+\frac{L_{ch}+\chi}{L}\right)\cdot
\left(1-\frac{\chi}{L}\right),\quad
\mbox{Pr}_{max}=\frac{1}{2}\left( 1+\frac{L_{ch}}{L}\right).
\end{equation}
Therefore, for a specified channel length the probability maximum in Eq.
(27)
is reached {\it at the interval boundary at $\chi=0$.}
If the channel length equals the effective state extent ($L_{ch}=L$),
this probability is unit.  In that case the eavesdropper can reliably
know each transmitted state and remain undetected. For
$L_{ch}=0$ the maximum is reached at $\chi=0$ and the probability is
1/2, which means that the eavesdropper simply guesses the transmitted
state.

Up to this moment, we have not yet taken into account the noise in
communication channel. Let us demonstrate now that in a noisy channel
the
probability (27) cannot exceed the corresponding value for the ideal
channel.
To do this, we shall take advantage of the relativistic causality
arguments.
The state modification induced by noise can be described by an
appropriate
instrument taking into account the relativistic restrictions imposed on
it.
The instrument can generally be written as [24--27]
\begin{equation}
{\cal T}[...]=\sum_k {\cal S}_k[...]{\cal S}_k^+,\quad
{\cal S}_k=\sqrt{\lambda_k}|\phi_k\rangle\langle\varphi_k|,\quad
\sum_k\lambda_k{\cal S}_k{\cal S}_k^+\le 1,\quad \lambda_k\ge 0,
\end{equation}
\begin{equation}
\mbox{Tr}\{ {\cal T} [|\psi_{0,1}\rangle\langle\psi_{0,1}|]
I(\Omega_E,t_e)\}=
\sum_k\mbox{Tr}\{ |\psi_{0,1}\rangle\langle\psi_{0,1}|
({\cal S}_k I(\Omega_E,t_e){\cal S}^+_k) \}\le
\sum_k\lambda_k\mbox{Tr}\{|\psi_{0,1}\rangle\langle\psi_{0,1}|
(|\varphi_k\rangle\langle\varphi_k|) \}\le
\end{equation}
\begin{displaymath}
\sum_k\lambda_k |\langle\varphi_k|\psi_{0,1}\rangle|^2\le
\sum_k \lambda_k \langle\varphi_k|\varphi_k\rangle
\langle\psi_{0,1}|\psi_{0,1}\rangle\le
\langle\psi_{0,1}|\psi_{0,1}\rangle=
\int_{\Omega_A}dx \Big|{\cal F}(x-t_A)\Big|^2=
\int_{\Omega_E}dx \Big|{\cal F}(x-t_E)\Big|^2.
\end{displaymath}
The last equality in Eq. (29) reflects the fact that the amplitude
${\cal F}(x-t_A)$ of state
\begin{equation}
|\psi_{0,1}\rangle=\int_{\Omega_A}dx {\cal F}(x-t_A)|x,t_A\rangle
\otimes |e_{0,1}\rangle=
\int_{\Omega_E}dx {\cal F}(x-t_E)|x,t_E\rangle\otimes |e_{0,1}\rangle
\end{equation}
at time $t_A$ is completely localized in the domain $x\in\Omega_A$,
and will not be completely localized in domain $x\in\Omega_E$
earlier than at $t_E=t_A+\mbox{dist}(\Omega_E,\Omega_A)$.
Strictly speaking, in the field theory the problem cannot be treated as
a
single-particle one in the sense that the operator $\hat{S}$ involves
the processes associated with creation of particles and absorption of
other photons entering the channel from the environment which can be
treated as a sort of noise. However, detection of these external photons
can obviously provide no additional information on the transmitted bit.
{\it Therefore, the probability for the eavesdropper to know an
individual
transmitted bit and remain undetected by the legitimate user B does not
exceed the corresponding probability $\mbox{Pr}_{max}=1/2(1+L_{ch}/L)$
for the ideal channel.}

Let us know describe the protocol.

\noindent
$\bullet$ At the time moments agreed upon beforehand, user A prepares
and sends into the communication channel the states
$|\psi_{0,1}\rangle$,
while user B performs measurements described by the identity resolution
(14,15). Only the bits which pass the test (i.e. those which produced
the outcomes in channels ${\cal P}_{0,1}(t_B)$) are kept, and all the
rest
transmissions are discarded (${\cal P}_{\bot}(t_B)$). If the channel
is ideal and eavesdropping is absent, each bit sent by user A is
reliably
identified by user B and contributes to the key (in contrast to the
cryptosystems based on non-orthogonal states).

\noindent
$\bullet$ Then the noise (identification error probability) is
estimated.
Users A and B disclose some of the transmissions and obtain an estimate
for the error probability $p_{err}$ counting the number of
inconsistencies
in their data sets. The disclosed bits are discarded from the
transmitted string.

\noindent
$\bullet$  The ``antedate'' coding is performed.
User A divides the remaining transmissions
into the groups each consisting of $k$ identical bits
(either all zeros or all units) and announces through a public channel
which transmission belongs to which group without disclosing the values
of the bits occurring in each group.
User B performs the error correction for each block employing a simple
majority voting principle [28]. The number $k$ is chosen sufficiently
large to reduce the error in identification of the block-wise bits
$\tilde{bit}(i)$ ($\tilde{0}=\overbrace{\{0,0,...0\}}^k$ and
$\tilde{1}=\overbrace{\{1,1,...1\}}^k$) below $\approx p_{err}^{k}\ll
p_{err}$. This procedure enhances the probability of survival of the
bit string finally accepted both users as the generated key. Then
the block-wise bits are assigned their reference numbers.

\noindent
$\bullet$ The users form $N+M$ parity bits
$Bit=\sum_{i=1}^{n}\oplus\tilde{bit}(i)$ from the block-wise bits.
This is achieved by user A announcing the reference numbers of the
block-wise bits included into each parity bit. Produced in this way
is a new string of parity bits.

\noindent
$\bullet$ Hashing procedure consisting of $M$ rounds is applied
to $N+M$ parity bits $Bit(j),\quad j=1..N+M$. To do this,
user A in each round chooses a random string $s_l$ of length
$N+M-l$ ($l=1...M$) and announces it to user B through a public
channel. Then user A and B check the parities of the subsets of bits in
their
strings ($Bit_A$ and $Bit_B$) comparing the parities with the string
$s_l$,
since $s_l\cdot Bit_A
=s_l\cdot Bit_B=(s_l\cdot Bit_A)\oplus (s_l \cdot Bit_B)= s_l\cdot
(Bit_A\oplus Bit_B)$. If the parities of the substrings coincide, one
bit
in the specified position is discarded form the strings $Bit_A$ and
$Bit_B$.
If the parities are different, the protocol is aborted. After $M$
successful
rounds the probability for the two remaining strings $Bit_A$ and $Bit_B$
possessed by users A and B, respectively, and each consisting of $N$
parity
bits to be different is [29]
\begin{equation}
\Pr\{s_A(N)\neq s_B(N)\}=2^{-M}.
\end{equation}
By appropriate choice of $M$ one can always make this probability
sufficiently small. Hence the first part of the protocol security
criterion (Eqs. (1) and (2)) is proved.

\noindent
$\bullet$ The probability for the eavesdropper to reliably know one of
the
parity bits and remain undetected is calculated in the following way.
The total number of ways in which every parity bit can be built from the
block-wise bits is [30]
\begin{equation}
\frac{1}{2}\sum_{i=0}^{n} C^{i\cdot k}_{n\cdot k}=
\frac{2^{n\cdot k}}{2k}\sum_{l=1}^{k} \cos^{n\cdot k}{(\frac{l\pi}{k})}
\cos{(nl\pi)} \approx\frac{1}{2k} 2^{n \cdot k}.
\end{equation}
The Hartley information of the set of block-wise strings is (to within
the rounding error) the number of binary symbols required to identify the
string
parity which practically coincides with the string length $n\cdot k$:
\begin{equation}
I=\mbox{log}_2(\frac{2^{n\cdot k}}{2k}\sum_{l=1}^{k} \cos^{n\cdot k}
{(\frac{l\pi}{k})}\cos{( nl\pi)} )\approx \eta \mbox{ } n\cdot k ,
\quad \eta \approx 1,
\end{equation}
that is almost all the bits in the string should be known. The
probability of
knowing every bit and passing the test is does not exceed (27) so that
the
conditional probability for the eavesdropper to know $N$ final bits
(key)
transmitted by user A and accepted by both users as the key is
(remember that $(1+L_{ch}/L)/2 < 1$)
\begin{equation}
\mbox{Pr}\{s_A(N)=s_E(N)\}=
2^{-N}\{1+ 2\cdot 2^{-\eta n\cdot k}[(1+L_{ch}/L)]
^{\eta\cdot n\cdot k}\}^N=2^{-N}(1+2\zeta)^N,\quad
\zeta=2^{-\eta n\cdot k}[(1+L_{ch}/L)]^{\eta\cdot n\cdot k}.
\end{equation}
Mutual information on the string of final bits of length $N$ possessed
by user A available to the eavesdropper is
\begin{equation}
I(A;E)=I(A)-I(A|E),\quad
I(A)=-\mbox{log}_2 2^{-N},\quad
I(A|E)=-\mbox{log}_2 \mbox{Pr}\{s_A(N)=s_E(N)\},
\end{equation}
where $I(A)$ is the proper information of the final string
$s_A(N)$ of length $N$, $I(A|E)$ is the conditional information
on the string $s_A(N)$ available to E.
Since all possible bit strings arise with the same probability and the
conditional probability for all strings is the same, we use the
information
which is not yet averaged over the string distribution. The conditional
information has a natural interpretation as the number of additional
bits
required for E to reliably recover the bit string of length $N$.
As to the mutual information, it is interpreted as the number of bits
measuring the information on string $s_A(N)$ of length $N$ available
to the eavesdropper [31].
Taking into account Eqs. (34,35) one obtains for the mutual information
between A and E:
\begin{equation}
I(A;E)=N-N + N\mbox{log}_2(1+2\zeta)\approx \frac{2N\cdot \zeta}{\ln
2}=2N\cdot
2^{-\eta n\cdot k}[(1+L_{ch}/L)]^{\eta\cdot n\cdot k}/{\ln 2}
\ll 1.
\end{equation}
For any specified $N$, $L_{ch}$, and $L$ ($L_{ch}<L$) this information
can be made exponentially small in the parameter $n\cdot k$.

\noindent
$\bullet$ Let us now show that the mutual information available to E
on the string $s_B(N)$ is also exponentially small.
Mutual information between A and B is
\begin{equation}
I(A;B)=I(A)-I(A|B)=-\mbox{log}_2 2^{-N} + \mbox{log}_2
\Pr\{s_A(N)=s_B(N)\}=N + \mbox{log}_2(1-2^{-M})\approx
N-\frac{2^{-M}}{\ln 2}.
\end{equation}
Taking advantage of the triangle inequality for the information [32]
one finally obtains
\begin{equation}
I(A|E)\le I(A|B)+I(B|E),\quad I(B|E)\ge N-(2N\cdot \zeta-2^{-M})/{\ln
2},\quad
I(B;E)\le (2N\cdot \zeta-2^{-M})/{\ln 2}\ll 1.
\end{equation}
Thus the second part of (Eqs. (3) and (4)) of the security criterion is
also proved.

We are grateful to Lev Vaidman for useful discussion of obtained
results.

This work was supported by the Russian Foundation for Basic Research
(grant \# 99-02-18127).

\end{document}